\documentclass[12pt]{article}
\usepackage{amssymb,amsmath,amsfonts}
\usepackage{epsfig}

\newcommand{\be}{\begin{equation}}
\newcommand{\ee}{\end{equation}}

\newcounter{mysubequation}[equation]

\newcommand{\TeV}{\,\mathrm{TeV}}

\newcommand{\half}{\frac{1}{2}}

\newcommand{\LL}{\mathcal{L}}
\newcommand{\OO}{\mathcal{O}}
\newcommand{\RR}{\mathcal{R}}
\newcommand{\eff}{{\text{eff}}}

\newcommand{\al}{\alpha^\prime}
\newcommand{\del}{\partial}
\newcommand{\lsim}{\,\raise.3ex\hbox{$<$\kern-.75em\lower1ex\hbox{$\sim$}}\,}
\newcommand{\gsim}{\,\raise.3ex\hbox{$>$\kern-.75em\lower1ex\hbox{$\sim$}}\,}

\begin{document}

\baselineskip=18pt
\setcounter{footnote}{0}
\setcounter{figure}{0}
\setcounter{table}{0}

\begin{titlepage}
\begin{flushright}
SLAC-PUB-11016\\ SU-ITP-05/08\\ hep-th/0507205
\end{flushright}
\vspace{.3in}
\begin{center}
{\Large \bf  N-flation}

\vspace{0.5cm}

{\bf\mbox{ S. Dimopoulos$^1$, S. Kachru$^{1,2}$,
J. McGreevy$^1$ and J.G. Wacker$^1$}}

\vspace{.5cm}

{\it $^1$ Physics Department, Stanford University, \\ Stanford,
California 94305, USA}

{\it $^2$ SLAC, Stanford University, \\ Menlo Park, California 94309, USA}

\end{center}
\vspace{.8cm}

\begin{abstract}
The presence of many axion fields
in four-dimensional string vacua
can lead to a simple,
radiatively stable realization
of chaotic inflation.
\end{abstract}

\bigskip
\bigskip


\end{titlepage}

\section{Introduction}

Slow roll inflation  \cite{inflation, chaotic}
is the leading candidate for early universe cosmology.
However,  finding a fully realistic model of inflation without fine-tuning
is an ongoing endeavor \cite{attempts}.
In this note we present a simple module for slow roll
inflation that appears to be  common in known
string compactifications.

The essential idea of this paper is that the inflaton is not any single field,
but a collection of $N$ fields.  Any one of the fields would not slow
roll for an appreciable number of e-foldings, but when taken together,
these fields self-damp and can slow roll for many e-foldings\footnote{
There are several examples of multi-field inflationary
models in the literature \cite{assisted, otherN}.
In particular, the `Assisted Inflation' model \cite{assisted}
exploits a similar idea using a specific exponential potential.
In each of the models of \cite{assisted, otherN}, issues of
radiative stability, $N$-scaling
and UV sensitivity need to be addressed.
}.
The predictions of the scenario are almost identical to those of
the original
$m^2 \phi^2$ model of chaotic inflation.

Verifying that this model of slow roll inflation
is under radiative control and not tuned requires detailed knowledge of
the low energy effective action of string theory, including higher order
curvature terms in the action, gravitational loop corrections, and
an understanding of moduli stabilization.   These details are important
for two reasons.
First, chaotic inflation
is robust from the low energy point of view, but with reasonable
assumptions about the ultraviolet dynamics,
a functional fine-tuning of the potential
is necessary to obtain a significant number of e-foldings.
We will show how this model evades these arguments.  Second,
we will show that while classically it appears as though
inflation can last for a period of time which is
parametrically long as a function of $N$,
radiative corrections change this
parametric scaling into a
numerical success.  This sets an upper limit to the number
of e-foldings achievable without tuning.
Therefore we will need to understand the sizes of radiative
corrections.

The organization of the paper is as follows.   In Sec.\ \ref{Sec: Pythagoras}
we describe the general idea of N-flation.  Here we make arguments about
the low energy effective theory and
identify what information we need from
the UV theory.  In Sec.\ \ref{Sec: Strings}  we discuss how N-flation
appears in a wide class of string theory compactifications.   In Sec.\ \ref{Sec: Repredictions}
we show that the inflationary predictions match those of standard $m^2\phi^2$ chaotic
inflation.

\section{Pythagoras Saves Chaotic Inflation}
\label{Sec: Pythagoras}

In this section we study a field theory with a potential of the form
\begin{eqnarray}
\label{ourpot}
V(\phi_n) = \sum_{n=1}^N V_n(\phi_n).
\end{eqnarray}
where each $V_n$ only depends on a single $\phi_n$.   Without the potential,
each $\phi_n$ would be  a Goldstone boson with independent shift symmetries
$\phi_n\rightarrow \phi_n + \delta_n$.  Each $V_n$ breaks a different shift symmetry,
in contrast to a general potential which would break all of the shift symmetries
in one shot.
We will take the potential to be periodic since the inflatons will ultimately be axions
\begin{eqnarray}
V_n(\phi_n) = \Lambda_n^4 \cos \left({2 \pi \phi_n \over f_n}\right) + \Lambda_n^{(2)\,4}
\cos \left({4 \pi \phi_n \over f_n}\right) + \cdots
\end{eqnarray}
where $f_n$ is the axion decay constant and $\Lambda_n$ is the dynamically
generated scale of the axion potential that typically arises from an instanton
expansion.  This scale can be many orders of magnitude
beneath the Planck scale.
Higher order instanton terms will give the higher harmonics in the potential,
but are of the size
\begin{eqnarray}
\Lambda^{(2)}_n \simeq \frac{\Lambda_n^2}{M}
 \end{eqnarray}
where $M$ is a UV scale.
If $\Lambda\ll M$ it is safe to drop all higher overtones of the potential.
Each $f_n$ will be less than the Planck scale, though not significantly so\footnote{
In the opposite regime $f > M_P$, which may not
be attainable in string theory \cite{banks},
one could make a model of `Natural Inflation' \cite{freese}.}.
Multi-instanton corrections to the potential can also violate the form of the
potential in Eq.\ \ref{ourpot}, leading to cross couplings between the axions
\begin{eqnarray}
\label{Eq: Cross Coupling}
V^{(2)}_{nm} = \frac{\Lambda_n^4 \Lambda_m^4}{M^4}
\cos \left({2\pi \phi_n \over f_n}\right) \cos \left( {2 \pi\phi_m \over f_m} \right) .
\end{eqnarray}
Thus, when we are in a regime where the potential in Eq.\ \ref{ourpot} is valid,
we drop higher order terms in the instanton expansion.
We will now show that a potential of this form can inflate.

For small field values the potential can be Taylor expanded
about its minimum as
\begin{eqnarray}
V_n (\phi_n) = \half m_n^2 \phi_n^2 - \frac{1}{24}\lambda_n \phi_n^4 +\cdots .
\end{eqnarray}
where $m_n=2\pi\Lambda_n^2/f_n$ and
$\lambda_n\simeq\left( 2\pi\Lambda_n /f_n \right)^4$.
For simplicity, we will take all masses $m_n =m$
in this main discussion.
In Sec.\ \ref{Sec: Cascading} we show
that examples with a spectrum of masses can still
inflate.

Consider an initial configuration where every axion field starts out
displaced from the minimum by $\langle\phi_{n 0}\rangle  = \alpha_n M_P$,
with the maximum displacement set by each axion decay constant
\begin{eqnarray}
\label{Eq: AlphaLimit}
\alpha_n^2 \lsim \frac{f_n^2}{M_P^2} .
\end{eqnarray}
Here we are tacitly assuming that we can hold each $\alpha_n$ fixed as we take
$N$ to be larger; we will address this issue in Sec.\ \ref{Sec: Strings}.
While each field has a
sub-Planckian vev, the total displacement from the origin
is super-Planckian, $\sim\sqrt{N}\alpha M_P$.
In polar coordinates, $\rho^2 \equiv \sum_n \phi_n^2$, the action has the form
\begin{eqnarray}
\LL \simeq
(\partial \rho)^2 + \rho^2( \partial\Omega)^2 - \half m^2 \rho^2
+ \frac{1}{24N} \lambda \rho^4 + \cdots
\end{eqnarray}
with $\langle\rho_0\rangle= \sqrt{N}\alpha M_P$.   The angular fields, $\Omega$,
have big kinetic terms from $\langle\rho^2\rangle\simeq N\alpha^2$,
and are easily over-damped
and drop out of inflationary dynamics.
The $N$ shift symmetries force corrections to the inflaton potential to be
subdominant in the large $N$ limit so that the potential can be trusted over a distance
$\sqrt{N} f > M_P$.
The form of the potential in Eq.\ \ref{ourpot} is crucially important for this to work;
if the potential were $SO(N)$ symmetric there would be no added control
of large vevs over a one-field model with many light (and irrelevant) fields, because there
would be $\OO(N^2)$ quartic couplings.
The quartic self-couplings are small and will be dropped from now on.
Finally, the volume of super-Planckian field space, $\rho > M_P$
grows much larger than sub-Planckian field space as the number of fields is increased.
This means that the typical initial condition
in the large $N$ limit is expected to be super-Planckian and suitable for chaotic inflation.

It is possible to use the radial variable for the inflaton.
Consequently,  the gross inflationary predictions of these models coincide with those of
$m^2\phi^2$ chaotic inflation.
Each $\phi_n$ field satisfies the equation of motion
\begin{equation}
\label{EOM}
\ddot \phi_n + 3H \dot \phi_n = - m^2 \phi_n
\end{equation}
with $3 H^2 = V/M_P^2= N \alpha^2 m^2$  and grows with the number of fields,
holding the initial condition of each field fixed.
Eq.\ \ref{EOM} shows that while each scalar feels the restoring force from its
own mass term, it feels the Hubble friction from the entire $N$-field
configuration.  For the initial condition $\phi_{n 0} = \alpha M_P$, the
theory inflates for
\begin{equation}
N_{e} = \frac{\alpha^2 N}{4}
\end{equation}
e-foldings until  $\langle\phi_n\rangle$ drops to $\sim M_P/\sqrt{N}$.

The slow roll parameters $\eta$ and $\epsilon$ are diagonal matrices
with each entry given by
\begin{equation}
\label{Eq: Slow Roll}
\eta \equiv M_P^2\frac{ V'' }{ V}
\sim \frac{1}{ \alpha^2 N},\hspace{0.2in} \epsilon
\equiv \frac{M_P^2 }{ 2} \left( \frac{ V'}{ V} \right)^2
\sim \frac{1}{ \alpha^2 N^2} .
\end{equation}

Density perturbations are given by
\begin{equation}
\label{ourcase}
\frac{\delta \rho }{ \rho} \sim N \alpha^2 \frac{m}{ M_P}
\end{equation}
Since $N\alpha^2$ sets the number of e-foldings, the small parameter controlling the
density perturbations $\frac{\delta \rho}{ \rho} \sim 2 \times 10^{-5}$
is the inflaton mass, requiring $m \sim 10^{10}$ TeV.

\subsection{Radiative Stability}
\label{Sec: Rad Cor}

In this subsection, we will study the radiative corrections to
the classical action in the previous section.
These corrections take several forms.
In turn, we will discuss the issues of large distances in field space,
renormalization of the axion potential,
the renormalization of Newton's constant, and finally the breaking
of global symmetries by small black holes.

There is a general worry that slow roll inflation over
Planckian field distances may not make sense in a quantum
theory of gravity.  The rough statement is that if
you go more than $\OO(M_P)$ away from a given minimum,
string scale modes can become light, and there is a different
effective theory with different degrees of freedom.   However,
axions are periodic (with periods smaller
than $M_P$) and we have functional control of the potential;
therefore we can safely consider all field values within the effective
field theory.

In order to get slow roll inflation, it was crucial that the cross couplings between
different axions were small.  One could worry that loop effects might
destabilize this form of the potential and spoil slow roll.  Each axion is endowed
with its own approximate shift symmetry.  In the low energy theory
only the potential breaks the shift symmetry.
This means  that any loop induced correction to the effective potential must
be proportional to the breaking and thus takes on the form
\begin{eqnarray}
\label{Eq: Eff Action}
\delta\LL_{\eff}(\phi_n) =
\sum_n b_nV_n''(\phi_n) \RR
+ \sum_{mn} \frac{c_{mn}}{M_P^4} V_n(\phi_n) V_m(\phi_m)+\cdots.
\end{eqnarray}
The first term in Eq.\ \ref{Eq: Eff Action} is the induced coupling to the Ricci
scalar which arises from one-loop gravity corrections.
Induced cross couplings are the same size as the multi-instanton in
Eq.\ \ref{Eq: Cross Coupling}
that we safely ignored earlier.
That these effects are sufficiently small not to spoil slow roll inflation
can be seen from the change in the slow-roll parameter
\begin{eqnarray}
\delta \eta \sim  (c_{mn} \eta+ b_n\varsigma) H^2/M_P^2
\end{eqnarray}
where $\varsigma \equiv M_P^4 V''''/V\sim N \eta^2$.
This analysis shows that there is not a ``low energy'' problem
with chaotic inflation.
In particular, corrections of the form $\phi_n^2 V/M_P^2$ are forbidden by the shift
symmetries of the axions.

A serious consideration is whether super-Planckian field configurations
have simply been swapped for a species problem, since a large number
of fields can enhance radiative corrections (for recent discussions
of similar issues see \cite{species}).
There is a quadratically divergent contribution to
the effective Planck mass from each light field
\begin{eqnarray}
\label{Eq: mplcorrection}
\delta M_P^2 \simeq \pm \frac{N}{16\pi^2} \Lambda_{{\rm UV}}^2
\end{eqnarray}
which can dilute gravity,
depending on the UV-sensitive sign.
Hence, the correction to the $\eta$ parameter which is induced by the
shift in $M_P$
\begin{equation}
\eta \simeq \frac{1}{N \alpha^2}\left( 1 \pm  \frac{N \Lambda_{\rm{UV}}^2}{16\pi^2 M_P^2}\right)
\end{equation}
dominates at very large $N$.
 This means that one can ${\it not}$ get a
parametrically large number of e-foldings
in a regime where the classical contribution
to the gravitational coupling is dominant.  There is a value of the number
of axions where the suppression of $\eta$ saturates
\begin{eqnarray}
\label{Eq: Max Efold}
N \simeq 16\pi^2 \frac{M_P^2}{\Lambda_{\text{UV}}^2}.
\end{eqnarray}
Substituting this into the expression for the number of e-foldings
\begin{eqnarray}
N_e^{\text{max}} \simeq 40 \alpha^2 \frac{M_P^2}{\Lambda_{\text{UV}}^2}.
\end{eqnarray}
This looks very promising, but is clearly UV sensitive.
We will address this in Sec.\ \ref{Sec: String Rad Cor}
where we estimate $\Lambda_{\rm{UV}}$
for the string realizations.
Whether the species problem is severe
enough to spoil slow roll inflation
is a detailed
numerical question.

The final worry is that small black holes violate  global symmetries
which include the shift symmetries of the axions.
These may generate unsuppressed potentials for the axions
and spoil slow roll inflation.  This will not be problematic in the string
realization because the axions' shift symmetries will descend from
short distance gauge symmetries which are not violated by black holes.

\subsection{Supersymmetric Radiative Stability}
\label{Sec: Susy Rad}

Supersymmetry is not crucial for this general inflationary mechanism --
the only required ingredient is that each field  is endowed with its own
softly broken shift symmetry.  If the low energy theory is supersymmetric,
then the arguments of the previous section need to be supplemented by those that
we consider here.  The string models we consider in the next section will
have low energy supersymmetry which introduces additional issues.
The axions each lie in their  own chiral superfield
\begin{eqnarray}
t_n =
\left( {\phi_n \over f_n} + i M^2\, R^2_n \right)
\end{eqnarray}
where $R^2_n$ is the modulus associated with $\phi_n$, and $M$ is a UV
scale.
For a supersymmetric theory, we need to use the supergravity effective potential
\begin{eqnarray}
V_{\text{sugra}}(\phi_n) = \exp\left(\frac{K}{M_P^2}\right)
\left(  \left|D W\right|^2 - 3 \frac{|W|^2}{M_P^2}\right).
\end{eqnarray}
There are three separate quantities in the potential
which might break the axion shift symmetries: $|D W|^2$,
$|W|^2$ and $K$.  The arguments of the previous section apply
directly to $|DW|^2$, but we will have to consider
$|W|^2$ and $K$ separately now.

The supergravity potential contains corrections to the
rigid supersymmetric
potential of the form $K/M_P^2$ and results in the supergravity
$\eta \sim \OO(1)$ problem \cite{susyeta}.  Because
our inflatons are axions, the K\"ahler potential is a function of
$ t_n - t_n^\dagger$, which is independent of the axion, $\phi_n$.
Therefore  the K\"ahler potential does not spoil slow roll inflation
\cite{yanagida}.  The K\"ahler potential
does give a mass for the moduli $R^2_n$ that causes them to roll down to
their respective minima quickly and decouple from inflationary dynamics.

The instanton induced superpotential is given by
\begin{eqnarray}
\label{Eq: Our W}
W \simeq W_0(S) +  \sum_n
w_n(S)~
e^{2 \pi i t_n} + \OO(e^{4\pi i t_n}).
\end{eqnarray}
with
\begin{eqnarray}
w_n(S) = w_{0n} + \mu^2_n S + \cdots
\end{eqnarray}
$W_0(S)$ parametrizes the physics which stabilizes the dilaton $S$;
its detailed form is irrelevant for our purposes and can be approximated by
\begin{eqnarray}
W_0(S) =  w_0 - m^2 S + \cdots.
\end{eqnarray}
The auxiliary field in the $S$ supermultiplet is given by
\begin{eqnarray}
F_S = -m^2 + \sum_n  \mu^2_n e^{2\pi i t_n}  + \OO(S, e^{4\pi i t_n}).
\end{eqnarray}
Solving $F_S=0$ (assuming $S$ is stabilized in a supersymmetric vacuum above
the Hubble scale of inflation), we find that the minimum is given by
\begin{eqnarray}
m^2 = \sum \mu^2_n \exp(2\pi i \langle t_n\rangle) .
\end{eqnarray}
Using this, the leading axion potential arises from the cross-term in 
the square of $F_S$
\begin{eqnarray}
\label{Eq: Susy Axion Potential}
|F_S|^2 &=&\nonumber
\left|
\sum_n\mu^2_n\Big( \exp(2\pi i \langle t_n\rangle)  - \exp( 2\pi i  t_n ) \Big) +\cdots
\right|^2 \\
\nonumber
= \sum_{n,m} e^{-2\pi M^2 (R_n^2 + R_m^2)} &\mu^2_n \mu^2_m 
&(1 - 2 \cos (2\pi \phi_m/f_m) + 
\cos(2\pi ( \phi_m/f_m -  \phi_n/f_n)) ) + \cdots
\end{eqnarray}
While the last term in the expression contains $N^2$ terms, each $\phi_n$ only
appears in $N$ terms, 
leading to the same parametric $N$ scaling that was
described in Sec. \ref{Sec: Pythagoras}.\footnote{
We thank L. McAllister for bringing this to our attention.  A more detailed
analysis of this class of models under various assumptions 
about the kinetic matrix for the axions and the distribution of the $R_n$s 
will appear in \cite{Liam}. 
}  Again, this is easiest to see in a radial variable. 
Let us 
consider any ray $|\phi_1|= |\phi_2| =\cdots=|\phi_N|\equiv \rho/\sqrt{N}$,
and for simplicity take all of the $\mu^2_n$ and $R_n^2$ to be the same.  
When averaged over the signs of the $\phi_n$, the potential in this approximation becomes
\begin{eqnarray}
V(\rho) \sim N^2 \exp(-4\pi M^2 R^2) \left( 1- 2 \cos 2\pi \rho/\sqrt{N} f + \cos 4\pi \rho/\sqrt{N} f \right).
\end{eqnarray}
Any higher order terms go as $\exp( - 6\pi M^2 R^2)$.   This has effectively halved
the value $f$ for a given $N$, but does not parametrically change the $N$ scalings.

The last term in the supergravity potential, $|W|^2$,
could introduce significant
cross couplings between the axions.   Using the form of the superpotential
in Eq.\ \ref{Eq: Our W}, we see that
all cross-couplings can be cast into the form of Eq. \ref{Eq: Susy Axion Potential}
and thus are higher-order gravity corrections to the potential already discussed.

\subsection{A Spectrum of Masses}
\label{Sec: Cascading}

So far we have considered the axions to have the same masses.
However we expect them to have a non-trivial spectrum.
In this section we briefly outline
the analysis for a more general set of masses.
We will show that N-flation is insensitive to this distribution.

A more realistic mass distribution is uniform
on a log scale.  For example, there could be hundreds  of (roughly degenerate)
fields in each decade of energy starting from near $M_P$ down to
$10^{10} \TeV$ or below.
This will result in sequential or multi-step inflationary periods, each one setting the stage
for the next.  If there are many fields at a sufficiently high density then we can approximate
the field index with a continuous label $\phi_n \rightarrow \phi(n)$ with masses
$m^2_n \rightarrow m^2(n)$.
For a uniform density of fields on a logarithmic energy scale, the masses take the form
\begin{eqnarray}
m^2(n) = M_P^2 e^{- n/\sigma}
\end{eqnarray}
where $\sigma$ is the density of fields per decade.
If the fields start with initial conditions $\phi(n,t=0)=\alpha M_P$,
then all of the fields are over-damped if $\sigma \alpha^2  \gg 1$.  At first only the heaviest fields
begin sliding down the potential.  After a Hubble time the 
heaviest fields are no longer over-damped.
Instead of immediately becoming under-damped and oscillating
(thereby ending inflation) they
remain critically damped due to the presence of the lighter fields.
Hubble changes slowly ($\dot{H}/H^2\sim 1/\sigma \alpha^2$)
so that the amount of time that a field of mass $m$ stays critically damped is
\begin{eqnarray}
\Delta t \sim \frac{H}{\dot{H}} = \frac{\sigma \alpha^2}{H} \sim \frac{\sigma\alpha^2}{m}.
\end{eqnarray}
During critical damping the fractional loss in amplitude is given by
\begin{eqnarray}
\frac{\phi_F}{\phi_I} \sim \exp(-m \Delta t) \sim \exp(-\sigma\alpha^2).
\end{eqnarray}
This shows that   all
the potential energy of the heaviest fields
is dissipated away before it can be converted
into kinetic energy.  Inflation proceeds until the final fields are no longer over-damped.
Schematically, the first period of inflation will create a large smooth patch (solving the patch problem),
the period 60 e-foldings from the end will give rise to $\delta\rho/\rho \sim 10^{-5}$, and the
last period will reheat the universe.

The general lesson is that if the axion masses
are densely spaced,
there will be sufficient self-damping to allow
inflation to proceed.

\section{The Many Axions of String Theory}
\label{Sec: Strings}

In oriented critical superstring theories,
there is a massless two-form
field, $B_{\mu\nu}$, and when the ten-dimensional theory is compactified to
four dimensions there are many independent two-cycles that $B_{\mu\nu}$
can wrap.  Each such cycle
results in an axion  at low energies.
Compactification on a six-manifold $M_6$ leads to $N = h^{(2)}(M_6)$ such axions;
$N$ can be very large\footnote{
It follows from known examples of
F-theory compactification that
there exist supersymmetric string models
with ${\cal O}(10^5)$ axions \cite{fourfolds}.}.
 These axions have independent
shift symmetries that keep them lighter than the scale of compactification
even in the absence of supersymmetry \cite{GSW}.  Ultraviolet physics can make
no contributions to the axion masses because above the scale of compactification,
these global symmetries become gauge symmetries of $B_{\mu\nu}$.
In particular, short distance physics ({\em e.g.} small black holes) cannot violate the shift symmetries of the axions.
In type II theories, similar considerations apply to the higher $p$-form fields.

The axions are paired into chiral superfields $t_n= \phi_n/f_n +i R^2_n/\alpha' $ where
$R_n^2$ is the modulus associated with the volume of the $n$th two-cycle.
The volume of $M_6$ is related to
the sizes of these cycles by
\begin{eqnarray}
\label{Eq: Volume}
V_6(t) = i \frac{\alpha^{\prime\,3}}{6}C^{lmn} t_l t_m t_n \Big|_{{\rm Re}~t =0}
\end{eqnarray}
where $C^{lmn}$ are determined solely by the topology of $M_6$, and
are integers.  The form $C$ is very sparse,
having only $\OO(N)$ nonzero entries in many models,
rather than
scaling as $\OO(N^3)$.
These integers may be negative; this can be interpreted
as resulting from a cycle that reduces the volume of the space as it grows larger.
First, notice that
if all of the intersection numbers are positive, then  the volume grows
as $N$  and $M^2_P$ falls as $1/N$, spoiling any gain in $\eta$ from having
many fields.    However, it is generic to have negative intersection numbers,
allowing string scale volumes despite the presence of many fields.  The
next question is whether this cancellation which makes the volume small
is just the tuning to get the potential sufficiently flat.  This is the
question of where it is natural to stabilize moduli.

At large radius, the K\"ahler potential for moduli is
$ \al K(t, \bar t) = - \ln \left( V_6(t-\bar t)/\al{}^3 \right)$,
with $V_6$  given in Eq.\ \ref{Eq: Volume}.
Therefore, the axion decay constants are
\begin{eqnarray}
\label{twothirdslaw}
\nonumber
 \frac{f_{mn}^2 }{M_P^2}
&=& \del_m \bar\del_n K = \frac{\alpha^{\prime~2}C^{mnl} R_l^2}{V_6}-
\frac{\alpha^{\prime~2} C^{mlk}R^2_l R^2_kC^{nl'k'}R^2_{l'} R^2_{k'}}{4V_6^2}
+ \OO(e^{- 4\pi R^2/\alpha'}) \\
&\sim& \frac{\alpha^{\prime\,2} R^2}{V_6}
\equiv \alpha^2.
\end{eqnarray}
where in the last approximation we have taken the volume to be slightly
larger than any individual radius (which we take to be approximately the same
size).
Note that the complicated second term in the first line of
Eq.\ \ref{twothirdslaw} only affects one of the $N$ eigenvalues of $f_{mn}$.
Where the volume enjoys cancellations between its various terms
while maintaining positivity of the metric on moduli space,
one can expect larger decay constants.
One important factor in determing where moduli are stabilized is
the volume of moduli space.  The volume of moduli space has measure
factor
\begin{eqnarray}
\det \frac{f^2_{mn}}{M_P^2}  \sim \left( \frac{\alpha^{\prime~2} R^2}{V_6}\right)^N.
\end{eqnarray}
Note therefore that where the measure is large, $f^2/M_P^2$ is also large.
This indicates that  on a significant portion of the  volume of the moduli space,
one finds string-scale volumes and $f^2\sim M_P^2$.
This is not the only factor in determining the distribution
of decay constants, but others are less well-understood and are model dependent.
Nevertheless, it is plausible that the moduli space volume form
favors small bulk volumes \cite{distributions} and hence large axion
decay constants.

The axions' shift symmetry is only broken by $W_n$ and not $K$, which respects
$t_n \rightarrow t_n + \delta_n$ in perturbation theory.
 The low energy effective action for
the supersymmetric theory is given by
\begin{eqnarray}
\LL=\int \!\!d^4\theta\; K\left(t-t^\dagger, S-S^\dagger\right)
+ \int \!\!d^2 \theta \left( \sum^{N}_n W_n(t_n, S) + W_0(S ) +\cdots \right)
+ \text{ h.c.}
\end{eqnarray}
The superpotential is
\begin{eqnarray}
W_n \simeq  \sum_\ell
w_n^\ell(S)~
e^{2 \pi i \ell t_n};
\end{eqnarray}
for axions associated with the NSNS B-field,
this is generated by worldsheet instantons.
This is the superpotential we studied in Sec.\ \ref{Sec: Susy Rad}.
There are also multi-instanton terms that can wrap two different
cycles and are given by
\begin{eqnarray}
W^{(2)}_{nm} \simeq
w_{nm}(S)~ e^{2 \pi  i( t_n+ t_m)}.
\end{eqnarray}
This is the same parametric scaling as described in Eq.\ \ref{Eq: Cross Coupling}.

A similar discussion applies to the axions arising from RR $p$-forms
in type II string theories,  where the role of the worldsheet instantons
is played by Euclidean D-branes.

\subsection{Radiative corrections in string theory}
\label{Sec: String Rad Cor}

In Sec.\ \ref{Sec: Rad Cor} we saw that we needed to estimate corrections
to $M_P^2$ that grew with the number of axions.  Any dynamics at distances
shorter than the compactification scale
cannot be sensitive to the number of axion fields
because they all descend from a single ten-dimensional field\footnote{For
concreteness and simplicity, the discussion here is appropriate for
heterotic Calabi-Yau compactification with the
gauge connection set equal to the spin connection.
A similar discussion would apply
to examples which are known to have worldsheet
instanton generated potentials,
{\it e.g.} \cite{Ftheoryduals}.}.   The easiest way
to proceed is to find corrections to the ten-dimensional action
which after
compactification become proportional to the number of fields.
These operators are higher derivative corrections to the gravitational effective action
that can arise both classically and at loop level.
The first term that becomes sensitive to the number of axions is
\cite{Rfour}
\begin{eqnarray}
\LL_{10D\;\eff}\simeq  M_*^8\left( \RR_{10} +
\zeta(3) {\al}^3 \RR^4_{10} +\cdots\right)
\end{eqnarray}
with  $M_*$ the 10D Planck scale: $M_*^{-8} = g_s^2 (\al)^4(2\pi)^7/2$.
Upon compactification, $\RR_{10}^4$ contains an
$(\RR_6\wedge \RR_6\wedge \RR_6)\RR_4$ term and
since
\begin{eqnarray}
 \int_{M_6} \RR_6\wedge \RR_6 \wedge \RR_6 = \frac{\chi(M_6)}{(2\pi)^3} ~,
\end{eqnarray}
integrating over the six internal dimensions gives
a correction to Newton's constant
proportional to $\chi $.
The Euler character
$\chi$ is a measure of the number of light species after compactification
\begin{eqnarray}
\chi(M_6)=2( N - \tilde{N})
\end{eqnarray}
where $\tilde{N}$ is the number of complex structure moduli of $M$.

The 4D effective Einstein-Hilbert term  for string theory compactified on a
Calabi-Yau six-manifold $M_6$ is of the form
\begin{eqnarray}
\label{fourD}
\LL_{4D\;\eff} = M_P^2\left(1 +  \chi(M_6)
\left( \frac{\al}{2\pi} \right)^3
\frac{\zeta(3)}{V_6}
\right) \RR_4
\end{eqnarray}
where $V_6$ is the volume of  the internal space.
The second term, which arises from the reduction of the
sigma-model four-loop $\RR^4$ correction \cite{Rfour},
is proportional to the
``density of cycles in string units."
It can be interpreted as
a back reaction of the internal space to packing
a huge amount of topology in a small volume.

There are higher order terms in the ten-dimensional effective action that are
suppressed by more powers of $\alpha^\prime/2\pi$.
These local operators, however, can never
scale more than linearly with the number of light species,
because the wavefunctions of these modes
are localized in the internal space.

There are also $g_s$ corrections to the effective action.
These loops can both renormalize the short distance
10D effective action and give terms that can only be
written in terms of operators in the 4D effective action.
The renormalization of the short distance effective action
({\it e.g.\!} to $\RR_{10}^4$)
is clearly suppressed by  $g_s^2$, and therefore is always subdominant
at weak enough string coupling \cite{antoniadis}.
IR contributions to the effective action are always cut off at the KK scale.
They can become sensitive to higher powers of the number of axions, but we
expect that it requires $n$ loops to become sensitive to $N^n$.
The string loop effects are also suppressed by at least $1/6\pi$
(by standard reasoning of naive dimensional analysis \cite{georgi}),  so the $n$ loop
contribution to $M_P^2$ could be as large as
\begin{eqnarray}
\left( \delta M_P^2\right)_n \sim \left(\frac{g_s^2 N}{6\pi}\right)^n M_{KK}^2 .
\end{eqnarray}
Therefore, if $g_s^2 \lsim 6\pi/N$,
string loop corrections can be safely ignored.

The leading effect on $M_P$ is given by
\begin{eqnarray}
\label{deltampl}
\frac{\delta M_P^2}{M_P^2} =   \frac{ \chi(M_6)}{8\pi^3}\zeta(3)  \frac{\al{}^3}{V_6}
\simeq \frac{ \chi(M_6)}{206}  \frac{\al}{R^2} \alpha^2
\end{eqnarray}
where we have used Eq.\ \ref{twothirdslaw} in the final expression.
The volume of moduli space is peaked around $V_6 \sim \al{}^3$ which
coincides with the largest values of $\alpha^2$.
Note that this formula cannot be trusted in a
regime where the correction to Newton's constant
cancels the tree-level term;
at small volumes, this occurs at $\chi(M_6) \lsim -200$.
Comparing our high- and low-energy estimates
for the renormalization of Newton's constant,
Eq.\ \ref{deltampl} and
Eq.\ \ref{Eq: mplcorrection},
we infer that
$\Lambda_{\rm UV}^2 \simeq M_P^2 \frac{ 2 \zeta(3)}{\pi}
\frac{ \al{}^3 }{ V_6} \frac{\chi }{  N}.$
The number of e-foldings is set by the number of axions, rather than
$\chi(M_6)$.  The number
of e-foldings (using Eq.\ \ref{Eq: Max Efold}
and setting $R^2/\al = 1$)
is then
\begin{eqnarray}
\label{badestimate}
N^{\text{max}}_e \simeq  \frac{2\pi^3}{\zeta(3)} \frac{N}{|\chi(M_6)|}\approx 26
\frac{1}{| 1- \tilde{N}/N|}.
\end{eqnarray}
Note that the initial value $\alpha$ cancels out of this expression.
Thus it takes a small cancellation between
the integers $N$ and $\tilde{N}$ to get 60 e-foldings.  For instance, there
is a Calabi-Yau with $(N,\tilde{N})= (251, 251)$ where the dominant correction
to
$M_P^2$ vanishes and it is possible to have $N_e \simeq N/4$.

A number of possibilities can help with the $\OO(1)$ factor.
In examples arising from type IIA string theory, the number of closed-string
axions is actually
$h^{(1,1)} + h^{(2,1)}$, namely $N+\tilde N$ in our notation.
Secondly,
in Eq.\ \ref{badestimate} we have conservatively
taken $R^2 \sim \al$;
however $R$ can be smaller, $R^2 \sim \al / 2\pi$,
while preserving a reasonable instanton expansion.
Next, there may be spaces for which the intersection form
allows $V_6$ large preserving the fact that many
decay constants satisfy $f/M_P \sim 1$.
Finally, it would be interesting to study the
large-$N$ statistics of Eq.\ \ref{twothirdslaw} on the space of CYs;
any robust large-$N$ scaling which shrinks more slowly
than $1/N$ would lead to a parametric win in the number of e-foldings.

N-flation required no model building or tuning of
continuous parameters to achieve slow roll inflation.
The number of axions needed for the requisite number of e-foldings
is (suggestively) at the high end of values available from
known string compactifications.

\section{Inflationary Repredictions}
\label{Sec: Repredictions}

In this section we quickly give the standard inflationary observables.
Throughout we will express our final answers in terms of $N_e$ (which
fixes $N \alpha^2$) and
$\delta \rho/\rho$ (which fixes $m$)
in order to demonstrate that the predictions are identical to those
in standard chaotic inflation.

In \cite{Sasaki}, we find a general formula for the tilt
which is applicable in this class of $N$-field models.
The tilt is
\begin{equation}
1-n = \frac{8}{ \alpha^2 N} = \frac{2}{N_e}.
\end{equation}
Here and below, one should set $\alpha^2 N \simeq 240$ to find the
values that would be relevant for the few e-foldings visible near
our present horizon.

The power in gravity waves \cite{Starobinsky} is (in the convention
of \cite{peiris})
\begin{equation}
P_{g} = \frac{2 }{ 3\pi^2} \frac{V }{ M_P^4} =
\frac{\alpha^2 N m^2}{3 \pi^2M_P^2} =\frac{4}{3\pi^2}\frac{1}{N_e} \left(\frac{\delta\rho}{\rho}\right)^2
\end{equation}
at the start of inflation where $\langle \phi_n \rangle \sim \alpha M_P$.

The spectral index of the gravitational waves is
\begin{equation}
n_g = 2\frac{\dot H }{ H^2} = -\frac{4}{ \alpha^2 N}= -\frac{1}{N_e}.
\end{equation}

The relative magnitude of the gravity waves
to density perturbations, $r$, is given by
\begin{equation}
\label{ourr}
r \sim \frac{P_{g}}{P_{\cal R}} \sim \frac{32}{ \alpha^2 N}= \frac{8}{N_e}.
\end{equation}

Non-Gaussian features in the spectrum of perturbations
remain a small effect.
To see this, we can use the formalism in \cite{Lyth}, where $f_{NL}$ is given by
\begin{eqnarray}
-3/5 f_{NL} =
{ \sum_{ij} N_i N_j N_{ij} \over 2 ( \sum_i N_i^2 )^2 }
+ \ln kL\; P/2 {\sum_i N_{ii}^3 \over (\sum_i N_i^2 )^3}  =   ( 1 + 6 ) \eta + \ln kL\;
\frac{P}{2} N \eta^3
 \end{eqnarray}
with $N_i = \partial N_e/\partial\phi_i$,
and $N_{ij} = \partial^2 N_e /\partial\phi_i  \partial \phi_j$,
and $P$ is the power spectrum in the inflaton.
The second term, while $N$ enhanced, is subdominant because of the additional
powers
of $m^2/M_P^2$.
This answer was to be expected from the similarity to
chaotic inflation.
The only difference from the calculation in {\em e.g.\ }\cite{riotto}
is that there are $N-1$ over-damped scalars  with
$m^2 \ll H^2$.
However, these additional fields have very small cross couplings and
are essentially free fields, and their relative contribution to the energy density (including
their quantum fluctuations) is down by at least $(H/M_P)^2$ compared to the inflaton.
In order for non-gaussianities to be visible, additional dynamics is needed.
It would be interesting to know under what
circumstances these effects would be visible.

Since the inflaton is a pseudo-Goldstone boson, the leading
couplings to other fields are at least dimension 5.  This lowers the
reheat temperature. For instance, a typical
coupling which respects the shift symmetry is
$\phi F_{\mu\nu} \tilde{F}^{\mu\nu}/M_P$.  This leads to an
inflaton decay width of
\begin{eqnarray}
\Gamma_\phi \sim \frac{m_\phi^3}{8\pi M_P^2}
\end{eqnarray}
and a reheat temperature of
\begin{eqnarray}
T_{\text{RH}} =\sqrt{\Gamma_\phi M_P} \simeq m_\phi \sqrt{\frac{m_\phi}{8\pi M_P}}.
\end{eqnarray}
For typical inflaton masses, $m_\phi \sim 10^{10}\TeV$, $T_{\text{RH}} = 10^7 \TeV$ and
light gravitationally coupled particles ({\it i.e.\!} gravitinos) are not reheated, eliminating the gravitino problem.

\vspace{0.6in}
\centerline{\bf{Acknowledgements}}
\vspace{0.2in}
We thank Nima Arkani-Hamed for collaboration at
various stages of this project.
We thank G. Dvali, R. Kallosh, A. Linde, L. McAllister,
and especially E. Silverstein for helpful discussions.
The research of SK was supported in part by a David and Lucile Packard
Foundation Fellowship for Science and Engineering,
and by the D.O.E. under contract DE-AC02-76SF00515.
SD, SK, JM and JW receive support from the
National Science Foundation under grant 0244728.

\end{document}